\documentclass[12pt]{article}
\usepackage{natbib}
\bibpunct{(}{)}{;}{a}{,}{,}
\usepackage{amsmath}

\begin{document}

% --- Author Metadata here ---
% --- End of Author Metadata ---

\title{The Nature of Novelty Detection\thanks {This study was supported by the Chinese National Key
Foundation Research \& Development Plan (2004CB318108) and Natural
Science Foundation (60223004, 60321002, 60303005). Special thanks
to Ellen Voorhees for suggestions about the organization and
presentation of the paper. }}

\author{
        Le Zhao\footnote {
       {State Key Lab of Intelligent Technologies and System;
       Dept. of Computer Science and Technology;
       Tsinghua University;
       Beijing 100084 P.R. China;
       zhaole@tsinghua.org.cn}
}
       \and Min Zhang \and Shaoping Ma\footnote {
       {State Key Lab of Intelligent Technologies and System;
       Dept. of Computer Science and Technology;
       Tsinghua University;
       Beijing 100084 P.R. China;
       \{z-m, msp\}@tsinghua.edu.cn}
} }
\maketitle

\renewcommand{\baselinestretch}{1.3}
\setlength{\baselineskip}{1.3\baselineskip}

\begin{abstract}
Sentence level novelty detection aims at reducing redundant
sentences from a sentence list. In the task, sentences appearing
later in the list with no new meanings are eliminated. Aiming at a
better accuracy for detecting redundancy, this paper reveals the
nature of the novelty detection task currently overlooked by the
Novelty community $-$ Novelty as a combination of the partial
overlap (PO, two sentences sharing common facts) and complete
overlap (CO, the first sentence covers all the facts of the second
sentence) relations. By formalizing novelty detection as a
combination of the two relations between sentences, new viewpoints
toward techniques dealing with Novelty are proposed. Among the
methods discussed, the similarity, overlap, pool and language
modeling approaches are commonly used. Furthermore, a novel
approach, selected pool method is provided, which is immediate
following the nature of the task. Experimental results obtained on
all the three currently available novelty datasets showed that
selected pool is significantly better or no worse than the current
methods. Knowledge about the nature of the task also affects the
evaluation methodologies. We propose new evaluation measures for
Novelty according to the nature of the task, as well as possible
directions for future study.
\end{abstract}


\section{Introduction}
From 2002 to 2004 \citep{harman02,ian03,ian04}, there were three
Novelty tracks held by the Text REtrieval Conference (TREC). The
focus was on sentence level query-specific (intra-topic) novelty
detection. In the tracks, first, sentences relevant to a given
topic (a query) are retrieved; secondly, according to the
chronological ordering of sentences, latter sentences which
provide no new meanings should be removed. Since the novelty task
is becoming increasingly important as information distribution and
users' needs for novel information from multiple sources increase,
it is drawing more and more attention from IR researchers all over
the world. In the many previous works, \citep{newsjunkie} provided
a vivid scenario showing the importance of novelty detection. The
novelty task can have applications in filtering \citep{yiz02},
question and answering (QA) or any other retrieval tasks that may
return redundancies to users.

Unlike many other natural language processing (NLP) tasks such as
retrieval, summarization, machine translation or QA, which mainly
deals with the relevance between documents and queries, or the
syntax or meanings of documents or sentences, novelty detection is
a task that deals with relations between sentences. Whether a
sentence's meanings are covered by another sentence or other
sentences is its major concern, while the meanings of sentences
themselves are indirectly involved.

In Novelty (by Novelty, we refer to the novelty detection task,
and so is for the rest of the paper), sentences whose entire
information content has already been returned by earlier retrieved
sentences should be eliminated; only novel sentences remain. In
this paper, the novelty or redundancy is Boolean valued; a
sentence is either redundant because of previous sentences or
novel (same as in the TREC Novelty tracks).

Most previous works \citep{yiz02, allan03, lzhao02, newsjunkie,
lzhao03} concentrated on the retrieval aspect of the novelty
detection task which treated Novelty as a single process of
retrieving novel sentences from a sentence list with possible
redundancies, and thus, inevitably overlooked the characteristics
of the novelty task we here propose. In this work, we review the
task from an ``overlap'' perspective. The overlap method is one
typical novelty measure introduced by \citep{lzhao02}. Later on,
the reader will see that overlap is not only a method, but also
one nature of Novelty. Note, the focus of this paper is on novelty
detection $-$ eliminating redundant sentences while preserving all
sentences that contain new information; whenever we refer to
``Novelty'' or ``the Novelty track'' in this paper, we are
referring to this step (namely, the ``task 2'' of the TREC 2003
and 2004 Novelty tracks). In ``task 1'' of the TREC Novelty
tracks, the participants first tried to retrieve the relevant
sentences from a collection of sentences, then to reduce the
redundancies of the retrieved sentences, while in task 2, the
relevant sentences were already given, and only redundancy
reductions were to be performed. The experimental results of this
paper were obtained on the Novelty 2003, 2004 task 2 datasets and
Yi Zhang's novelty collection, in which redundancy reductions were
performed on the sets of all the relevant and only the relevant
sentences (documents). The collections are consisted of about 50
topics, and of course with each topic a different set of relevant
sentences (documents). Though the units of novelty processing in
Yi Zhang's collection were documents, different from TREC's
sentences, our selected pool method was also proved effective on
this collection, which showed that our method generalizes well for
different units of processing. The qualities of the available
datasets could be another reason why researchers have not noticed
the nature of Novelty we here propose, there was only a short time
after all the three collections were established; more
importantly, in natural language processing tasks, theoretical
evidence from semantic theories though could provide a solid basis
for discussion, experiments done on more collections make people
more confident about the empirical validity of a proposed
framework: in this paper, the PO-CO relation formalization of the
novelty task.

In Yi Zhang's pioneering work on Novelty \citep{yiz02}, several
questions were raised regarding the redundancy measure and the
novelty detection procedure: symmetric or asymmetric redundancy
measure, sentence-to-sentence or sentence-to-multiple-sentences
comparison model. We hope that this study has found satisfying
answers to the questions both theoretically and empirically.
Problems such as novelty depending on the user (what's novel for
the user) are not the focus. Actually various forms of background
information can be brought into the novelty task. For example,
logic can be used to derive more meanings given a fixed set of
sentences. Physics, geographical information or even personal
knowledge of a user can also be adopted. Though only the original
meanings were considered in this study, we can still incorporate
into our framework further information or rules other than the
original lexical meanings, with certain modifications.

An outline for the rest of the paper is as follows: Section
\ref{chap:overlap} starting from the similarity method, summarizes
the widely used overlap methods and some current difficulties in
novelty computation. Section \ref{chap:2relation} is the heart of
the paper, in which the nature of the novelty detection task is
provided as a combination of the PO-CO relations. Though our
formalization can be seen as a direct derivation from the semantic
theories of natural language, this formalization is actually
independent of the representations of sentences or documents (as
sets of meanings, as sets of terms, or as language models). Some
implications toward techniques dealing with Novelty suggested by
this nature will also be discussed in this section, such as the
use of language modeling and clustering techniques in Novelty. In
section \ref{chap:selpool}, we try to address the current
difficulties in novelty computation empirically, with the help of
its nature, which leads to the selected pool method. We provide,
in section \ref{chap:exp}, the corresponding experimental results
on the three novelty detection datasets, which reveals the
comparative advantage of selected pool to the overlap and pool
methods. Furthermore, in section \ref{chap:eval}, following the
experiments in the previous sections, we discuss the evaluation
methodologies for Novelty from two distinct views, and propose a
more reasonable measure. Section \ref{chap:conclu} concludes the
paper and provides possible directions for novelty detection.

Here, we apologize in advance for the dispersed experimental
results among the sections: section \ref{chap:overlap} (which
contains an introduction to the three novelty datasets and
experiments for overlap and similarity method), section
\ref{chap:exp} (which contains the comparisons between selected
pool and other methods on the three collections; these comparisons
are essential in supporting the main thesis of the paper), and
section \ref{chap:eval} (which evaluates the results between
similarity and overlap in section \ref{chap:overlap} with a
different measure: SPSM). The experiments were distributed to the
corresponding sections to maintain a continuity of discussion in
this paper.

In the course of our investigation, meanings of sentences are
inevitably involved. Our mathematical treatment of meanings is
from a unique viewpoint, different from any previous semantic
theories; we studied the relations between sentences in their
meanings. Although these treatments are far from complete, they do
work for Novelty at least. Nevertheless, novelty detection remains
a difficult task which is determined by the complexities and
arbitrariness of natural language. The purpose of our study was to
identify exactly where the difficulties lie, and divide the
difficulties into as many small pieces as possible. The field of
novelty detection must still face problems that cannot be easily
solved, relating to natural language understanding.

\section{The overlap method}\label{chap:overlap}
In the previous works, there were always two standard themes of
novelty detection techniques. In one theme, to judge the current
sentence, first, redundancy comparisons between the current
sentence and each of the previous sentences were performed. Next,
the maximum of the redundancy scores obtained from the first step
was compared against a threshold ($\alpha$) to finally decide
whether the current sentence is redundant; if the maximum
redundancy score exceeded $\alpha$, the current sentence would be
classified as redundant. Simple similarity method \citep{yiz02}
and overlap method \citep{lzhao02} both adopted this one-to-one
comparison paradigm. In the other theme, the redundancy score
between the current sentence and the pool of all the previous
sentences together was used against threshold $\alpha$ to make the
redundancy decision. The simple pool \citep{lzhao02} and the
interpolated aggregate smoothing Language model \citep{allan03}
applied this paradigm.

The two themes were adopted because of the conception of the
novelty detection task that in judging a current sentence, all
previous sentences should be used. Next, we will show that this
conception is generally wrong, because novelty judgment is not
what we used to think a single inseparable judgment process.

We introduce three implementations of the above two themes related
to this investigation: the similarity method, simple term overlap
method and simple pool method. Improvements of our method over the
methods will shown in the experiments section.

\subsection{From similarity to overlap}
There are two notions essential in Novelty: the
\emph{differentiation of meanings} and the \emph{chronological
ordering} of acquisition of knowledge; if a new sentence contains
meanings that are \emph{different} from any other
\emph{previously} known meanings (facts), it is novel. (For
example, ``Tom is five'' is different from ``Tom has a sister''
even though Tom appears in both.) Consequently, if humans were
unable to differentiate the meanings of sentences, the novelty
task would no longer exist.

From the \emph{differentiation of meanings} we know that if the
meaning of a sentence is the same as some known fact, the sentence
is redundant. So a symmetric similarity measure between
\emph{sentences} can be used to estimate the symmetric ``same''
relation between \emph{meanings}. A sentence sufficiently similar
to a previous one is considered redundant. Taking a first look at
the novelty task, anyone would probably come up with a similarity
measure. The differentiation of meanings is probably the only
reasonable explanation for the use of symmetric methods in
Novelty, which depends on identifying meaning (fact) with
sentence; one sentence could only contain one fact.

Although similarity has been proven to be effective experimentally
\citep{yiz02}, \citep{lzhao03}, if we think twice, when one
sentence's meanings are covered by another, this relation is not
necessarily symmetric, because sentences may contain multiple
meanings. An asymmetric overlap measure should be used eventually
(\citep{yiz02} and \citep{lzhao02} mentioned such belief). The
overlap method in \citep{lzhao02} was proved to be stable among
different data collections \citep{lzhao02, lzhao03, lzhao04}. The
similarity and overlap methods presented in this paper were
defined as in \citep{lzhao02, lzhao03}:
\begin{equation}
\begin{split}
Sim(A, B)=\frac{|A\cap B|}{|A\cup B|}
=\frac{\sum_{i \in A\cap B}{\min(A_{i},B_{i})}}{\sum_{i \in A\cup B}{\max(A_{i},B_{i})}}\\
Overlap B_{A}=\frac{|A\cap B|}{|B|}=\frac{\sum_{i \in A\cap
B}{\min(A_{i},B_{i})}}{\sum_{i \in B}{B_{i}}}
\end{split}
\end{equation}

$Overlap B_{A}$ is the overlap of sentence B by a previous
sentence A. A$_{i}$ is the TFIDF weight of term i in A
\citep{salton88}. In experiments, thresholds were set to judge
whether two sentences are sufficiently similar, or whether a large
enough portion of one sentence is overlapped by another.

Surprisingly, despite the theoretical advantage of overlap,
similarity is empirically better than asymmetric methods like the
overlap method, as experimental results from \citep{yiz02} and
\citep{lzhao02} indicated.

Before we present the experimental results, let us first introduce
the three datasets of novelty detection.

In Zhang Yi's pioneering work on large scale empirical study of
the novelty detection problem \citep{yiz02}, a document level
novelty detection dataset (nvyiz) was constructed on the archive
of news articles from Associated Press (AP) year 1988 to 1990 and
Wall Street Journal (WSJ) 1988 to 1990. This collection has
totally 50 topics, but 5 of them lacks human redundancy
assessments which were excluded from the experiments in this
paper. In \citep{yiz02}, two notions of redundancy were used in
the assessments: absolutely redundant and somewhat redundant. In
the experiments below, we are concerned only with the notion of
absolute redundance which is the same as from the TREC Novelty
collections.

In TREC 2003 and 2004, two datasets TREC Novelty 2003 (nv03) and
2004 (nv04) were constructed, also on newswire articles. Both
consist of 50 topics, but use sentences as units of processing
instead of documents (sentences from 25 relevant articles for each
topic were used to construct nv03 and nv04).

Experiments in this paper were performed on these three
collections, the only public text collections currently available
for novelty detection research.

\begin{table}
\centering \caption{Similarity and the overlap method}
\label{tab:sim}
\begin{tabular}{|c|c|c|c|c|c|} \hline
\textbf{nv04} 5 docs&\#ret&Av.P&Av.R&Av.F&\#novel\\ \hline

s0.4&986&0.688&0.977&0.790&627\\ \hline

o0.7&974&0.694&0.964&0.786&634\\ \hline

\textbf{nv04} 25docs&\#ret&Av.P&Av.R&Av.F&\#novel\\ \hline

s0.4&7008&0.463&0.957&0.610&3282\\ \hline

o0.7&6965&0.462&0.950&0.608&3255\\ \hline

\textbf{nv03}&\#ret&Av.P&Av.R&Av.F&\#novel\\ \hline

s0.4&13495&0.719&0.978&0.817&9962\\ \hline

o0.7&13303&0.719&0.972&0.815&9836\\ \hline

\textbf{nvyiz} &\#ret&Av.P&Av.R&Av.F&\#novel\\ \hline

s0.4&9082&0.919&0.977&0.946&8313\\ \hline

o0.8&9349&0.909&0.988&0.945&8452\\ \hline

\end{tabular}
\end{table}

Table \ref{tab:sim} provides for each run: \#ret - the total
number of sentences for the 50 topics returned by a run (judged to
be novel by a run), Av.P - precision of the true novel sentences
in the returned averaged over 50 topics, Av.R - average recall of
novel sentences, Av.F - average F-measure (F-measure trades off
between precision and recall), and \#novel - number of novel
sentences returned. In all the tables, we used the following
abbreviations: ``s $\alpha$'' for similarity with threshold
$\alpha$, ``o $\alpha$'' for overlap with threshold $\alpha$, and
``p $\alpha$'' for pool. In Table \ref{tab:sim}, ``o0.7'' is the
overlap method with threshold $\alpha = 0.7$; ``s0.4'' is the
similarity method with $\alpha = 0.4$. (both overlap and
similarity thresholds were chosen to optimal on the test
collection.)

From the table, we can see that in F-measure, similarity was
uniformly better than overlap on the three collections: nv03
(0.817 vs. 0.815, significant at p=0.04 by a paired t-test), nv04
(for all the 25 documents, 0.610 vs. 0.607 but not significant by
paired t-test, for the first 5 documents, 0.790 vs. 0.786, not
significant; we keep the results from the first 5 documents for
nv04 due to reasons that will be given in section
\ref{chap:eval}), and nvyiz (0.946 vs. 0.945, similarity is
better, but not significant). The F-measure difference was small
because overlap and similarity differ slightly.

\subsection{The pool method}
In the above subsection, only sentence to sentence comparison is
considered. But for general novelty detection, since all ``old''
sentences should be used to judge the current sentence, a possible
method would be to compare the current sentence with all previous
sentences. A pool method would be an obvious choice, in which
overlap between the pool of terms from all previous sentences and
the set of terms from the current sentence is computed, with a
fixed threshold $\alpha$ for redundancy judgment like in overlap.

But features like TFIDF weighted terms, being only surface
features of sentences not the exact meanings, make this pool
consisting of all previous sentences too noisy to perform well.
Section \ref{chap:exp} compares the performance of overlap and
pool method on 2003, 2004 Novelty datasets and Yi Zhang's novelty
collection; in Table \ref{tab:per}, an overall comparison is
given. On one hand, overlap returned about 30\% redundant
sentences (in table \ref{tab:sim}, average precision is about
70\%), which suggests to remove more sentences, on the other hand,
the pool method which removes more sentences is too ``noisy'', we
resolve this difficulty by looking deeper into the task.

\section{The two relations}\label{chap:2relation}
By ``relation'' we mean the mathematical relation; a relation $R$
between the elements of a set $A$ is a subset $C$ of the Cartesian
product $A \times A$. Any $a \in A$ and $b \in A$, $aRb$, if $(a,
b)\in C$. In this paper, $A$ is a set of sentences, and we deal
with relations between sentences.

In which follows, mathematics will help us clarify the conception
about Novelty. In the novelty task, one relation requires much
attention: the relation whether a sentence is overlapped by
another or some others. But the definition of this relation is
never clearly stated in any previous work about Novelty, because
this relation is not exactly one relation! Previous literature,
such as \citep{yiz02, newsjunkie, lzhao02, lzhao03} have mixed the
following two relations with very different properties in forming
this ``one relation'' in Novelty. (\citep{cmu02} only considered
the $>_{co}$ relation between two sentences.)

\subsection{The two relations}
First, the partial order relation $>_{co}$, we called the
\emph{complete overlap relation}. One sentence A $>_{co}$ B, if A
contains all the meanings of sentence B. This relation is a
partial order relation. It is transitive and antisymmetric. For
sentences A, B and C:

\begin{enumerate}
\item A $>_{co}$ A (Reflexivity).

\item If A $>_{co}$ B and B $>_{co}$ A, then A = B in meaning
(Antisymmetry).

\item If A $>_{co}$ B and B $>_{co}$ C, then A $>_{co}$ C
(Transitivity).
\end{enumerate}

In Yi's work \citep{yiz02}, only the third property is presented
explicitly as an assumption. The above three properties together
characterize the complete overlap (CO) relation.

Second, the symmetric relation $>_{po}$, we called \emph{partial
overlap relation}. A $>_{po}$ B, if A and B have meanings in
common. Note that having common meanings does not require A to
completely overlap B, though complete overlap is sufficient for
partial overlap. This relation is non-transitive and symmetric.
For sentences A, B and C:

\begin{enumerate}
\item A $>_{po}$ A (Reflexivity).

\item If A $>_{po}$ B then B $>_{po}$ A (Symmetry).

\item If A $>_{po}$ B and B $>_{po}$ C, A and C need not have the
$>_{po}$ relation. (E.g., A = \{a\}, B = \{a, b\}, C = \{b\}.
Here, A $>_{po}$ B and B $>_{po}$ C, but A C do not have this PO
relation. No transitivity here).

\item If A $>_{po}$ B then $\exists\;C \neq \emptyset$ such that A
$>_{co}$ C and B $>_{co}$ C (\emph{Separation of meanings}). Here,
C is a separated sentence containing common meanings of A and B,
but need not contain all the common meanings of A and B.

\item If A $>_{co}$ B, $B \neq \emptyset$ then A $>_{po}$ B
(Complete overlapping is sufficient for partial overlapping).

\item If A $>_{co}$ B and B $>_{po}$ C, then A $>_{po}$ C. Here, A
is called a CO expansion of B, and this property states that CO
expansions preserve PO relations.

\item If A $>_{po}$ B and B $>_{co}$ C, A and C need not have the
$>_{po}$ relation.

\item If A $>_{po}$ B then $\exists \;C = A \cap B, C \neq
\emptyset$ (\emph{The intersection definition}). Here, C contains
all the common meanings.
\end{enumerate}

As the PO relation is symmetric, we called the sentences that are
PO related to one sentence its PO relatives. (e.g., for sentence A
in \{A: A $>_{po}$ B and A $>_{po}$ C\}, B and C are called A's PO
relatives, and similarly A is also B's and C's PO relative.)

In the above properties, (1, 2, 3, 5, 6 and 7) are the basic
properties of the PO relation as they can be derived from having
common-meaning definition, or from property (4) $-$ separation of
meanings alone, or property (8) alone. Property (4) is
sufficiently strong for the PO relation, but may not be necessary.
Property (8) is even sufficient for (4).

In the case of multiple sentences (e.g., A, B and C) overlapping
one single sentence (say D), the PO relation conforms to reality,
because to have an overlap relation, A, B and C must all be D's PO
relative (not necessarily $>_{co}$ D each), and together $A \cup B
\cup C >_{co} D$.

Note that for sentences we assume there are also operations and
relation like in set theory: $\cup$ $\cap$ and $\subset$, but they
need not be exactly the same as in set theory. In the
introduction, we said we can adopt background information and
rules such as geographical information and logic rules. The only
difference it will make is that there should be corresponding
modifications to the operations of the sentences. For example, $A
\cup B$ will be $A \cup B \cup$ \{the facts derived from A, B and
background information (if there is any) according to the rules\}.

In PO relations which are actually symmetric, we still use the
``$>$'' sign usually used to denote asymmetric relations. The
reason for this is that even though the PO relation itself is
symmetric, in the novelty task, where sentences are aligned along
a time line and only previous sentences can overlap a subsequent
one, thus, an asymmetry is imposed onto the PO relation. If A
$>_{po}$ B and B $>_{po}$ A, then A and B must appear at the same
time, which means A and B must be the same sentence. Hence we
could see clearly that the asymmetry of the PO relation is
external; it should not be mixed up with other properties of the
PO relation (unfortunately there was already a paper that did make
the mistake).

Note that the relations defined above are completely different
from the ``partially redundant'' and ``absolutely redundant'' in
\citep{yiz02}, where the redundancy is more subjective, as being
judged by assessors. The two relations we here defined are more
objective; they are Boolean valued, and the CO relation must be
either completely redundant or novel, which is closer to the
notion of ``absolutely redundant'', thus only the ``absolutely
redundant'' judgments in Yi Zhang's collection were used for the
experiments of this paper.

\subsection{Sets of facts}
In this subsection, we provide an explanation of the PO-CO
framework with semantic theories of language. Facts (which can be
represented as logical expressions) are meanings of the statements
that can be asserted as either true or false. In Novelty, only
sentences that tell clear and complete facts are considered. Even
though Tom appears in both ``Tom is five'' and ``Tom has a
sister'', the two statements are about two distinct facts, and
therefore, there are no common meanings between the two.
Throughout the paper, we are talking about ``meanings'' of
sentences, to be exact, it is actually the senses of sentences; we
distinguish reference and sense from the ambiguous word
``meaning'' like \citeauthor{gamut} \citeyearpar{gamut} did.
Novelty requires senses, not references, because it is intensional
in its nature rather than extensional, since it asks the question:
"Is the sentence novel?" rather than "Is it true?". Actually, the
above relational structure and properties of the relations arise
from the discrepancy between the units of novelty processing (i.e.
sentences) and the units of novelty definition (i.e. sense $-$
what is novel is actually individual senses, not sentences;
sentence is a much larger unit).

From the discussions of the previous sections, it may seem that
meanings of sentences are actually treated as sets of facts, or
similar to sets. We have even used sets in the examples. The set
of facts assumption for meanings is strong enough to provide all
previous properties listed. Since the set assumption is very
strong (which may even be the finest we can attain), we can use
sets to provide counter examples, as in the PO relation property
(3). But using sets introduces a problem; the set definition is
too strong, and has a narrower range of application. We should
generalize it little by little to the weakest assumptions we can
possibly achieve.

When we define the PO relation, there are actually three different
definitions. For \emph{the first}: (A B) is a PO pair if there is
common meaning between them. This definition is precise in the
sense that there are no assumptions about what meanings of
sentences would be like. In this definition, the properties (1, 2,
3, 5, 6 and 7) of the PO relation can be derived. In spite of its
simplicity, this PO definition is too ambiguous and should be
formalized to bring the PO relation into the PO-CO framework. This
can be achieved by \emph{the second definition}, which defines the
PO relation using the three properties of the CO relation and the
\emph{separation of meanings} (A $>_{po}$ B if $\exists \;C \neq
\emptyset$ such that A $>_{co}$ C and B $>_{co}$ C). Separation of
meanings is stronger than the first definition, because it says if
there are common meanings, some common meanings can be separated
(from A and B to a sentence C). \emph{The third definition} is the
intersection definition, which requires that for A $>_{po}$ B,
there exists a maximum sentence C = $A \cap B$. Here, maximum
means for any sentence S, if A $>_{co}$ S and B $>_{co}$ S then C
$>_{co}$ S. This definition is stronger than the separation of
meanings definition, since the separation of meanings can be
derived from it. None of the three definitions require meanings to
be treated as sets. The set assumption is even stronger than the
intersection definition. That is to say, in our definitions of CO
and PO relations, meanings of sentences need not be exact sets of
facts.

Every assumption here has its exceptions, of course. But at least
it's likely that in most cases the weak assumptions are not far
from the reality or from the users' needs if we just focus our
interest on the novelty detection in news stories where only
simple facts and events are involved. (Cases such as abnormal
state detection in industrial plant monitoring or novelty
detection in robot navigation \citep{saunders01} apparently seem
quite different from the text novelty task we are considering
here; even natural language is not involved. For these
applications, usually a deviancy measure for a new event to the
current probabilistic model alone is enough. But there can be
similar improvements like those we have brought into sentence
level novelty detection: introducing a PO relation and locating
the PO relatives before using deviancy measure. This means, the
PO-CO framework is a true nature of novelty detection, and exists
in every novelty detection task, not necessarily text novelty
detection.) What definition we shall choose at a specific occasion
depends on what properties we need in processing, but if we adopt
stronger and finer properties like separation of meanings or
intersection property or even set assumption, we must be aware
that the results we attain can only be applied to more limited
cases.

Here are some examples. A: ``Tom has a sister; she is reading a
book'', B: ``Tom's sister is reading a book''.

Sentence A contains more meaning than B because A states that Tom
has a sister (if we take only lexical information into account,
implications or presumptions of sentences are not considered). A
consists of two facts, but B only one. So A $>_{co}$ B. This
example can still be explained under sets of facts assumption, but
is surely less obvious than in ``Tom is five, and Tom goes to
school'' $>_{po}$ ``Tom is a five-year-old boy'' where the common
meaning can be separated as ``Tom is five''. The following will be
even more obscure.

C: ``I frightened the cat, and it ran away'', D: ``I frightened
the cat, so it ran away''.

C contains only two facts, but D contains two facts the same as C,
and also a belief that the cat ran away because I frightened it.
So D $>_{co}$ C. If sentences become more complicated, even for
the most sophisticated minds, it will be a difficult task to count
the facts in them. This is especially true when we consider more
background information or implications of meanings, because not
only can sentences have generated meanings but also there may be
contradictions derived from original sentences or the meanings
implied by them. Even if the assumptions do not fail, it is still
difficult to program a computer to solve them.

If, for example, we take emotional facts implied by sentences into
account, it will be difficult for the separation of meanings
assumption to hold: E: ``You savagely killed the cat'', F: ``You
murdered the cat''. The separation of the emotional subtleties
between E and F seems difficult. As we consider more facets of the
natural language, since the emotional suggestions of the terms
like ``murder'' or ``savagely'' are hardly exact and clear,
probabilistic models or fuzzy models may be of use.

Here in defining CO and PO relations, we only set up assumptions
about relations between sentences, since this is the least the
novelty task requires. The meanings of a sentence, whether
behaving like a set or not, are not necessarily concerned. At
least, we are very fortunate, as whatever definition among the
three we adopt, we can always have the several basic properties of
the PO relation.

Interestingly, there is also a correspondence \citep{fmri99} of
the PO and CO relations in the human brain during novelty
processing to the ``\emph{retrieval of related semantic
concepts}'' at the right prefrontal cortex (which usually actively
maintains context information during performance of working memory
tasks) and ``\emph{registration of deviancy}'' at the superior
temporal gyrus (the language and music processing center), which
strongly supports the discussions here. No matter how a PO
relation is defined or what the meanings of sentences may look
like, there should be a procedure that corresponds to the
\emph{retrieval of related semantic concepts}. This correspondence
is the best evidence that the PO relation actually widely exists,
and thus should not be neglected in novelty processing. But as we
saw in the above examples, if we are to practically use these
relations, there are many factors to be defined and specified
(such as what background information or rule to use, whether
implications or presumptions of sentences are considered and
setting up rules to resolve contradictions in the data, etc.), to
resolve uncertainties and rule out difficult cases in the natural
language.

\subsection{Some direct results from the relations}
After clarifying the nature of the novelty task, we can have some
nontrivial examples (applications) explained under the framework
of PO-CO relations.

A first example to see will be a method for Novelty that uses
clustering techniques \citep{lzhao03} (the Subtopic III method:
sentences are clustered into several classes and only sentences
within one class can have an overlap relation; overlaps between
clusters are not considered). As we know from the properties of
the PO relation, PO relations actually differ in one point from
equivalent relations: transitivity. PO relations are not
transitive, thus there can be no equivalent classes. The usage of
the clustering methods in the novelty task has an intrinsic
difficulty - the sentences need not necessarily form classes. So
introducing clustering techniques without taking this fact into
account can be harmful. In TREC 2003, the Subtopic III method was
shown to be ineffective. (The work \citep{yyang02} is different
from the intra-topic clustering discussed here. In
\citep{yyang02}, inter-topic clustering of documents were
performed, which is not our concern.)

A second example is the uses of language models (LM) in novelty
detection. There can be two usages of LM. In the first, like in
retrieval \citep{ponte98}, a generation probability of the current
sentence on the basis of previous sentences can be used to
estimate redundancy. Take for example, the task of ranking new
documents according to novelty given a known set of seed documents
\citep{newsjunkie}. According to the PO-CO framework, each
different newly appearing document, the LM for the previous
sentences should be constructed on the PO relatives of the new
document. The document sets used to construct the models could be
different for different new documents; thus, the comparison of
generation probability of the two new sentences using two
respective LMs is not mathematically justified. This is clearer
under the measure theoretic view of probability. The two different
models impose two distinct measures onto the event space. In
ranking documents in generation probability, what we are doing is
measuring two objects (the two new documents) using two different
rulers (the two models). This explains the intuition that if two
facts (A and B) are different and both are novel, it is impossible
to judge whether A is more novel than B or not. Since Novelty
requires only the differentiation of meanings, the ranking of
documents here must have been imposed by attributes other than
novelty (such as the amount of new information or the number of
new meanings). In practice (\citep{yiz02, newsjunkie, allan03}),
another usage of LM is common; for a current document, two LMs are
constructed for previous documents and the current document
respectively, and the KL-divergence between the two models is used
to approximate the degree of novelty. This use of LM, unlike
generation probability, is mathematically justified. However,
there is no step of finding PO relatives; all previous documents
are used. Because of this, it can be easily adopted into the PO-CO
framework by constructing the LM on the PO relatives.

Next, there is an important and direct implementation that
benefits from the successful distinguishing of the above PO-CO
relations.

\subsection{Novelty $-$ a complex task}\label{sec:complexNV}
Now, we return to the novelty task itself. Once we are clear about
the two relations discussed above, we can see immediately that the
novelty task we used to refer to as one single task can be
considered as actually consisting of two independent subtasks.

The first step is to find out the pairs of sentences that share
common meanings. (For a current sentence, this step is just
locating the previous sentences having PO relation with the
current one.) In this subtask, the \emph{separation of meanings}
definition can be useful, as in determining whether a pair has PO
relation, we only need to separate some common meaning. This
subtask can have its own judgment and evaluation method. One
apparent suggestion would be the error rate of the PO pairs (with
false alarms and misses as errors).

The next step is to judge whether a current sentence is completely
overlapped by previous PO relatives, with all the known PO pairs.
We may need to separate all common meanings between two PO
sentences in this subtask. And combining all the common meanings
of the previous sentences with respect to the current one, we will
finally be able to judge whether all meanings of the current
sentence are covered by the previous sentences.

In practice, if all the sentences are short, containing only one
simple fact, there is no need to use the PO relation; one CO step
would serve the purpose well, and there will be almost no
difference between the asymmetric overlap and the symmetric
similarity measure. The longer the sentences are, the more likely
multiple facts exist in a single sentence, and the more likely
methods that adopt the PO-CO framework will work better than the
methods that treat the complex Novelty as one single task. In the
real world data, informative articles always try to include
several facts in one single sentence (usually, with the aid of
clauses), which justifies applying the PO-CO framework to real
world data.

But the two subtasks are still difficult in the sense that they
have to deal with complicated cases, outliers of the simplified
assumptions we proposed. Even within the scope of the assumptions,
a computational solution to manipulate facts in the novelty task
is still not apparent. But since we have broken down the novelty
task into two subtasks where problems and difficulties are fewer
than the complex task that takes Novelty as a whole, it is
expectable that novelty research will move a step forward.

We are now able to see that the problems mentioned in the
introduction (symmetric or asymmetric novelty measure, one-to-one
or multiple-to-one comparison) arose because of an unclear
perception of the novelty task, and these questions are gone once
we take the view from the nature of the novelty task. But this
viewpoint still cannot explain how these questions arose
empirically. And our study of novelty, described below, tried to
investigate the empirical facet of the questions.

\section{The selected pool method}\label{chap:selpool}
Following the previous section, the final and best unit for
processing novelty would seem to be facts (i.e. logical forms
formalized from sentences). Unfortunately, without proper
formalizations, computers do not know what a fact is. And the task
of changing natural language sentences into logical forms is still
too difficult. When we do novelty detection, we have to use units
such as documents or sentences to base our novelty computation on.
Therefore, the two classification steps (PO: the step of
classifying whether two sentences are PO related, and CO:
classification of whether PO relatives of a sentence $>_{co}$ the
current sentence) always exist. Here we use the word
classification in the sense as in Pattern Classification, by
\citeauthor{duda} \citeyearpar{duda}.

From the stance of the PO-CO framework, it is clear that the
previous overlap and pool methods came about because of the
ambiguous conception of the novelty task that when making
\emph{the novelty judgment} of the current sentence, we could and
should use all the previous sentences in the list, while as a
matter of fact, all the previous sentences should be used in the
PO judgment, not necessarily the CO step. Accordingly, we propose
a selected pool method, in which only sentences that are related
to the current sentence are included in the pool (the PO step),
followed by a pool-sentence overlap judgment (the CO step). In the
experiments, if the TFIDF overlap score of the current sentence by
a previous sentence exceeded the selection threshold $\beta$, that
previous sentence was considered to be PO related to the current
sentence. By setting the threshold $\beta$ to be 0, we include all
previous sentences in the pool - the selected pool turns back into
the simple pool method. Setting $\beta$ to be the threshold for
pool-sentence overlap judgment $\alpha$, the selected pool becomes
the simple overlap method. Table \ref{tab:selpool} shows the
change in the performance of selected pool as $\beta$ changes. A
comparison between selected pool, and overlap and pool methods,
with automatically learned parameters $\alpha$ and $\beta$ using
cross validation will be provided in the next section, which is
summarized as Table \ref{tab:per}. What we must point out is that
the overlap score as a feature for making PO and CO decisions is
very coarse and certainly not perfect, which suggests possible
further improvements.

\section{Experiments and analyses}\label{chap:exp}
We provide experiments comparing selected pool and the baseline
methods (overlap and simple pool) on Novelty 2003 (nv03), 2004
(nv04) and Yi Zhang's novelty collection (nvyiz). Analyses
concerning the different characteristics of the three collections
and the differences in the performance of the different methods
will be also be stated here.

\begin{table}
\centering \caption{The relationship between overlap, pool and
selected pool methods with a changing parameter:
$\beta$}\label{tab:selpool}
\begin{tabular}{|c|c|c|c|c|l|} \hline
\textbf{nv04}&\#ret&Av.P&Av.R&Av.F&Difference\\ \hline
p0.7&5713&0.495&0.864&0.615&192 novel in\\ \cline{1-5}
sp0.7s2.0&6205&0.487&0.911&0.620&492 extra\\ \hline
sp0.7s2.0&6205&0.487&0.911&0.620&80 novel in\\ \cline{1-5}
sp0.7s3.0&6552&0.475&0.929&0.615&347 extra\\ \hline
sp0.7s3.0&6552&0.475&0.929&0.615&56 novel in\\ \cline{1-5}
sp0.7s4.0&6818&0.466&0.942&0.609&266 extra\\ \hline
sp0.7s4.0&6818&0.466&0.942&0.609&15 novel in\\ \cline{1-5}
sp0.7s5.0&6893&0.464&0.945&0.608&75 extra\\ \hline
sp0.7s5.0&6893&0.464&0.945&0.608&22 novel in\\ \cline{1-5}
o0.7&6965&0.462&0.950&0.608&72 extra\\ \hline
\textbf{nv03}&\#ret&Av.P&Av.R&Av.F&Difference\\ \hline
p0.7&9127&0.755&0.762&0.744&2763 novel\\ \cline{1-5}
sp0.7s5.0&13250&0.720&0.969&0.815&4123 extra\\ \hline
sp0.7s5.0&13250&0.720&0.969&0.815&25 novel\\ \cline{1-5}
o0.7&13303&0.719&0.972&0.815&53 extra\\ \hline
\textbf{nvyiz}&\#ret&Av.P&Av.R&Av.F&Difference\\ \hline
sp0.8s5.0&9199&0.914&0.978&0.943&70 novel\\ \cline{1-5}
sp0.8s6.0&9296&0.911&0.985&0.946&97 extra\\ \hline
sp0.8s6.0&9296&0.911&0.985&0.946&19 novel\\ \cline{1-5}
o0.8&9349&0.909&0.988&0.945&53 extra\\ \hline
\end{tabular}
\end{table}

In Table \ref{tab:selpool}, ``sp$\alpha$ s$\beta$'' is an
abbreviation for selected pool method with overlap threshold
$\alpha$ and selection threshold $\beta$. To avoid confusion
between the two thresholds, the selection threshold was defined to
be 8 times the actual one-to-one overlap threshold, so o0.7 =
sp0.7s5.6, with selection threshold 5.6/8=0.7. As the parameter
for selection changes, the method changes gradually from pool to
overlap. The selected pool with a higher selection threshold will
include fewer sentences in the pool, and thus will return more
sentences than with a lower selection threshold. The last column
of the table (``Difference'') is the number of additional returned
novel sentences in the totality of the extra sentences returned.
In F-measure s2.0 is better than p0.7, and s5.0 is almost the same
as o0.7. But for the additional returned sentences, only a small
part (about 1/3 to 1/4) of the additional returned sentences were
novel. Simple derivation showed that to increase the F-measure of
a set of results, additionally returning a set with precision
higher than P$\div$(P+R) is sufficient, where P and R are the
precision and recall of the original result set. For example, if
P=0.5 and R=0.9, including a set with precision greater than 0.36
already increases F-measure. This strange property of the
F-measure can be misleading when comparing different Novelty
methods only using the F-measure (e.g. similarity and overlap in
Table \ref{tab:sim}).

\subsection{Within-collection analyses}
On nv04 collection, the F-measure for sp0.7s2.0 (the best
performing selected pool) was significantly better than that for
o0.7 (the baseline overlap method of the best selected pool) by a
paired t-test (0.620 vs. 0.608, significant at p = 0.000006; of
all the 50 topics, 40 increased, 9 decreased, 1 remained the
same); if we consider \#errors made in novelty judgments, the
change from sp0.7s2.0 to o0.7 is more conspicuous (44 topics
decreased in \#errors $-$ improved, 3 increased $-$ degraded, 3
remained unchanged, with an average improvement of 8.5\%). On the
nvyiz collection, sp0.8s6.0 was slightly better than o0.8 in
average F-measure (0.946 vs. 0.945, p=0.30, not significant by a
t-test. But in \#errors, 15 topics improved, 7 degraded, 23 did
not change, sp0.8s6.0 was significantly better than o0.8 at
p=0.037). On the nv03 collection, sp0.7s5.0 was not significantly
better than o0.7, and was also no worse. These experiments on the
three collections suggested that multiple-to-one comparison is no
worse and sometimes better than one-to-one comparison if we use a
proper method like selected pool.

Now we are able to answer the question mentioned in the
introduction: one-to-one or multiple-to-one comparison. In
\citep{yiz02}, the multiple-to-one comparison was actually an
all-to-one comparison, like in the simple pool method, and simple
pool was significantly worse than overlap on nvyiz and nv03 (0.744
vs. 0.815, significant at p=0.0000000001) collections, but was
significantly better than overlap on nv04 (0.615 vs. 0.608,
significant at p=0.05), which suggests that the pool method is
unstable among datasets. The multiple-to-one comparison in
\citep{yiz02} was worse than one-to-one because the authors of
that paper were not clear about the PO-CO framework we here
propose and thus were unable to find an adequate way of using the
multiple-to-one comparison theme, whereas novelty computation
should use a multiple-to-one theme rather than one-to-one
comparison.

For the nv04 collection, the best performance of the selected pool
(sp0.7s2.0) was observed around the top runs submitted to TREC
2004 task 2 (Best run: City U. Dublin average F-measure: 0.622,
second: Meiji F: 0.619), among all runs with language modeling
approaches, cosine similarity measure, information gain and named
entity recognition \citep{ian04}. Even the worst (simple overlap)
could be ranked as high as 7$^{th}$. Thus, the overall performance
of the selected pool method as a technique that adopts the PO-CO
framework is encouraging.

\subsection{Inter-collection experiments and analyses}
Above are analyses within collections; inter-collection
comparisons of the collections themselves and of the performance
differences of the methods on different collections are provided
below:

Although nv03 and nv04 are datasets selected from the same set of
topics and from the same newswire data collection, the redundancy
rate by human assessments in nv03 is 34.1\% while in nv04 53.7\%.
This difference was surprising but unexplained by \citep{ian04}.
We believe that this difference in human assessed redundancy rate
is the cause for the difference in performance of selected pool on
the nv03 and nv04 collections. Selected pool is better on nv04
than nv03 probably because of the possible different
characteristics in the Novelty 2003 and 2004 human assessments -
04 contains more multiple-to-one overlapping cases while
one-to-one dominates in 03. There is no direct evidence for this
conjecture (human assessments are somewhat incomplete for nv03 and
nv04 collections), but it seems to be the most probable
explanation for the different behaviors of selected pool. It is
possible that with a much shorter list of relevant sentences for
each topic, (the rate of relevant sentences is almost less than
half that of nv03, this allows assessors to consider
multiple-to-one overlap cases more easily) when they were
constructing the nv04 dataset, the assessors paid more attention
to the multiple-to-one overlapping cases. This also is a feasible
explanation for the higher redundancy percentage in nv04 than that
of nv03.

Compared to the nv03 and nv04 collections, nvyiz has a more
complete structure (for each redundant document, the human
assessments also include all the previous documents that actually
make this document redundant). Therefore we can have direct
evidence from the human assessments showing that nvyiz has a per
topic redundancy rate of 10.8\%, multiple-to-one cases occupy
about 34.7\% of the 10.8\% redundancies. Because of the existence
of those multiple-to-one cases, the selected pool performed better
than simple overlap on nvyiz collection.

One last thing about the comparison between overlap and selected
pool is how to choose the parameters $\alpha$ and $\beta$. As
selected pool has one degree more freedom than overlap $-$
parameter $\beta$, does selected pool tend to overfit because of
its superior learning ability? To answer this question, we did
Leave-One-Out (LOO) estimations to estimate the expected F-measure
and expected \#errors of overlap and selected pool. In these
experiments, for each topic, the other 49 topics were used for
training, the one topic left out was used for validation. Because
the parameters were few, the entire parameter space was searched
at the training step. Paired t-tests on F-measures of the 50 (45
for nvyiz) test topics showed that on nvyiz selected pool was
better than overlap in F-measure (0.946 vs. 0.945, but not
significant, p = 0.30); on nv04 selected pool was significantly
better than overlap (0.620 vs. 0.614, significant at p=0.036); on
nv03 selected pool and overlap performed almost the same (0.815
vs. 0.815, overlap was slightly better, but not significant). The
important thing here is that the LOO estimate of selected pool
performed almost the same as the selected pool with the best
parameter setting, which means selected pool is stable in spite of
its greater learning ability.

The performance of selected pool and simple pool compared to
overlap on the three (nv03 nv04 and nvyiz) collections are
summarized in table \ref{tab:per}. In the table, ``$--$'' stands
for significantly worse than overlap on the corresponding
collection; ``++'' stands for significantly better under both
F-measure and \#errors than the overlap method; ``+'' stands for
not significant improvement in average F, but significant
improvement under \#errors; ``0'' stands for almost no difference.

\begin{table}
\centering \caption{Performance compared to
overlap}\label{tab:per}
\begin{tabular}{|c|c|c|c|}
  \hline
  Collections: & \textbf{nv03} & \textbf{nv04} & \textbf{nvyiz} \\
  \hline
  Simple pool & $--$ & ++ & $--$ \\ \hline
  Selected pool & 0 & ++ & + \\
  \hline
\end{tabular}
\end{table}

In the next section about evaluation, we delve into further
measuring of novelty techniques and propose measures suggested by
the PO-CO framework. In Yi's work \citep{yiz02}, precision and
recall for redundancy and number of mistakes in novelty judgments
were used. Based on the framework of PO-CO relations, the measures
we investigated were finer and could reveal richer contents.

\section{Evaluation methodologies}\label{chap:eval}
There can be two different viewpoints to the novelty task.

\emph{a. Novelty task as retrieval of novel sentences}

A set of novel sentences is to be retrieved from a stream of
sentences that may contain redundancies, and in the novelty
judgment for a current sentence, all previous sentences (they are
the acquired knowledge) can be used. This viewpoint is taken by
the TREC Novelty tracks \citep{harman02}, and is seemingly
intuitive, because this is just what the novelty task aims at. In
this study, the measure for evaluating results from this viewpoint
was designated as the Standard Novelty Measure (SNM), which is the
F-measure for the novel sentences:
\begin{equation}SNM=\frac{2 \times P \times R}{P+R}\end{equation}

P is the precision of the novel sentences returned, and R is
recall. As we will see later, some things are missed when only the
SNM is used for judging the results.

\emph{b. Novelty task $-$ a closer look}

When asked why one sentence is redundant, one must be able to
point out the exact previous sentences that make the current
sentence redundant (its PO relatives). So for PO-CO based methods
there should be a finer evaluation method that involves a closer
look, which does not only take into account the number of
correctly retrieved novel sentences but also the previous
sentences covering the redundant sentences. Even if a judgment for
the novelty of the current sentence is correct, if the previous
sentences that cover this current sentence are not correctly
judged, this chance output of the system certainly cannot be
called correct. We must construct a new kind of judgment that can
prevent the above fallacy.

In our method, we treat the novelty task as a classification of
overlapping sentence pairs. Measuring from a classification
viewpoint is what the previous TREC Novelty tracks lacked. In our
evaluation method, the error rate of the classification is used,
as in the following Pairwise Sentence Measure (PSM):
\begin{equation}\begin{split}
PSM=&1-\frac{\#\mathrm{misclassified\, pairs(system\; output)}}{\mathrm{\#total\,pairs(judgment)}}\\
\mathrm{where,}\;\;&\#\mathrm{misclassified\,pairs}\\
=&\mathrm{\#missed\,pairs}+\mathrm{\#false\,overlapping\,pairs.}\\
\end{split}\end{equation}

Here, by ``\#'' we mean ``the number of''. In the second formula,
the missed pairs and the false overlapping pairs are usually
called misses and false alarms in the classification terminology
(face detection for example), whereas the task here is to detect
the overlapping pairs. We define the PSM more clearly; for a run
A, suppose $A_{R}$ is the set of redundant sentences judged by A,
and $A_{N}$ the novel sentences judged by run A (the sentences
finally returned by A). Immediately, $A_{R} \coprod A_{N} = C$
(disjoint union), where $C$ is the collection of all sentences.
$i$ is a sentence, $\Re$ is the set of redundant sentences by
judgment (true redundant sentences), $PO_{i}$ is the set of true
PO relatives of sentence $i$, and $SPO_{i}$ is the set of
sentences run A judges as the PO relatives of $i$. For any X set
of sentences, suppose, $|X|$ is the usual measure of $X$ (the
number of elements in $X$), $|X|_{R}$ is the redundancy measure of
$X$ (the number of redundant ones in $X$), and $|X|_{N}$ is the
novelty measure of $X$ (the number of novel ones in $X$). Here, we
have $|X|_{N} + |X|_{R} = |X|$, $|X|_{R} = |X \cap \Re|$.

$\mathrm{\#misclassified \,pairs}$ and $\mathrm{\#total\,pairs}$
are further defined in equation (\ref{eq:furtherPSM}):
\begin{equation}
    \begin{split}\label{eq:furtherPSM}
\#\mathrm{misclassified}=&\sum\nolimits_{i \in A_{R}-\Re}{|SPO_{i}|}+\sum\nolimits_{i \in \Re-A_{R}}{|PO_{i}|}\\
+\sum\nolimits_{i \in A_{R}\cap\Re}&{|(SPO_{i}-PO_{i})\cup(PO_{i}-SPO_{i})|}\\
\#\mathrm{total\,pairs}\;\;\;=&\sum\nolimits_{i \in \Re}{|PO_{i}|}
    \end{split}
\end{equation}

With more correct pairs judged, the PSM will be larger, just as
the SNM increases when more novel sentences are found. Even
better, the PSM is a finer measure compared to the SNM. The SNM
does not correspond to the number of correctly classified pairs,
but, if the correct pairs are known, the precision and recall for
the novel (or redundant) sentences can be derived from them
directly. If the human assessments for PSM are known, we should
restrict the use of SNM and adopt this better choice.

In the experiments, unfortunately, as we were unable to determine
all the correct pairs for PSM evaluation for all the 50 topics;
there are thousands of sentences and tens of thousands of sentence
pairs in the Novelty datasets (quite a tremendous task for the
limited labor force available). Also, not all methods can return
the PO relatives of a sentence, so we adopted an alternative that
only made use of the human assessments currently available.

\emph{Simplified PSM (SPSM)}: since the $PO$ and $SPO$ are
unknown, we define $|PO_{i}|$ and $|SPO_{i}|$ to behave like
indicator functions, i.e.
\begin{displaymath}\begin{split}
\sum\nolimits_{i \in A_{R}-\Re}{|SPO_{i}|}=|A_{R}-\Re |,\;\sum\nolimits_{i \in A_{R}-\Re}{|PO_{i}|}=0\\
\sum\nolimits_{i \in
\Re-A_{R}}{|PO_{i}|}=|\Re-A_{R}|,\;\sum\nolimits_{i \in
\Re-A_{R}}{|SPO_{i}|}=0
\end{split}\end{displaymath}

Then, for a run A,
\begin{displaymath}\begin{split}
&SPSM(A)\\
=&(|\Re|-|A_{R}-\Re|-|\Re-A_{R}|)/|\Re|\\
=&(-|A_{R}|+|A_{R}\cap\Re|+|A_{R}\cap\Re|)/|\Re|\\
=&(|A_{N}|-|C|+2|A_{R}|_{R})/|\Re|\\
 &(since\; |A_{R}|_{R}=|\Re|-|A_{N}|_{R}=|\Re|-|A_{N}|+|A_{N}|_{N})\\
=&(2|\Re|-|A_{N}|+2|A_{N}|_{N}-|C|)/|\Re|\\
=&(|A_{N}|_{N}-|A_{N}|_{R}-(|C|-2|\Re|))/|\Re|\\
\end{split}\end{displaymath}

This simplification is intuitively proper; SPSM increases as more
novel sentences are returned, and decreases with more redundancies
returned. Also, the SPSM is linearly related to the
Redundancy-Mistake measure in \citep{yiz02} - an increase in SPSM
always corresponds to a decrease in the Redundancy-Mistake, if for
one topic (for an average over all the topics, SPSM does not
always equal to Redundancy-Mistakes, because $|\Re|$ could be
different for different topics). If a run A returns more novel
sentences on the basis of B (i.e. $A_{N} \supset B_{N}$),
$|A_{N}|_{N}-|B_{N}|_{N}=|A_{N}-B_{N}|_{N}$, then
\begin{displaymath}SPSM(A)-SPSM(B)=(|A_{N}-B_{N}|_{N}-|A_{N}-B_{N}|_{R})/|\Re|\end{displaymath}

\#novel in $A_{N} - B_{N}$ (number of novel sentences in the
additional returned sentences) corresponds to the decrease in
\#false alarms, and \#redundant in $A_{N} - B_{N}$ corresponds to
the increase in \#misses. This is shown in the last column of
Table \ref{tab:selpool}. Under SPSM, in Table \ref{tab:selpool},
selected pool (sp0.7s5.0) on nv03 dataset is slightly better than
overlap, and simple pool is far worse than both selected pool and
overlap. For nv04 the improvement of selected pool is consistent
under both SNM and SPSM. For nvyiz under SPSM, none of the
differences between similarity, overlap and selected pool is
significant.

\begin{table}
\centering \caption{Comparisons between similarity and overlap
under different measures}\label{tab:SPSM}
\begin{tabular}{|c|c|c|c|} \hline
\textbf{nv04} 5 docs&SNM&Mistake\%&SPSM\\ \hline

s0.4&0.790&29.5\%&0.1094\\ \hline

o0.7&0.786&30.2\%&0.1196\\ \hline

\textbf{nv04} all 25 docs&SNM&Mistake\%&SPSM\\ \hline

s0.4&0.610&47.8\%&0.1607\\ \hline

o0.7&0.608&47.9\%&0.1561\\ \hline

\textbf{nv03} 5 docs&SNM&Mistake\%&SPSM\\ \hline

s0.4&0.872&19.9\%&0.2105\\ \hline

o0.7&0.872&19.5\%&0.1970\\ \hline

\textbf{nv03} all 25 docs&SNM&Mistake\%&SPSM\\ \hline

s0.4&0.817&26.4\%&0.2645\\ \hline

o0.7&0.815&26.7\%&0.2320\\ \hline

\textbf{nvyiz}&SNM&Mistake\%&SPSM\\ \hline

s0.4&0.946&9.74\%&0.0773\\ \hline

o0.8&0.945&10.02\%&0.0236\\ \hline
\end{tabular}
\end{table}

Here, in Table \ref{tab:SPSM} we compare the similarity method
with the overlap method for nv04 nv03 and nvyiz datasets under
SPSM. The results indicate that under SPSM, the overlap method is
comparable to the similarity method. Similarity is not always
better than overlap; overlap is better on the collections of the
first 5 documents of nv04; the differences between SPSMs of
similarity and overlap are all insignificant by paired t-tests,
even though the change in SPSM on the nvyiz collection is
relatively large: 0.0237 vs. 0.0773. That is to say, similarity is
better than overlap on all the three datasets under SNM because of
the use of SNM. (The asymmetric novelty methods in \citep{yiz02}
were far less effective than similarity, and those asymmetric
methods were less effective than both the similarity and the
overlap method in this study.) The reason we keep the results from
the first 5 documents of the nv03 and nv04 collections is that,
with increasing number of documents, the possibility of
multiple-to-one overlap cases will also increase, which could
affect the evaluation of simple overlap and similarity methods
which only consider one-to-one overlapping cases. The better
performance of similarity on nv03 and nvyiz datasets suggests that
these two collections possibly contain more near duplicates, while
nv04 contains more overlapping cases (not just simple duplicates)
between sentences. Till this point we have both theoretically and
empirically answered the questions proposed by \citep{yiz02}:
symmetric or asymmetric measure, one-to-one or multiple-to-one
comparison.

As an alternative to PSM, the Simplified PSM is economical; it is
close to PSM, while requiring only information about the novel
sentences. Unlike PSM, we do not have to reassess the TREC Novelty
datasets before computing the SPSM.

\section{Conclusions and future work}\label{chap:conclu}
The PO-CO framework and related discussions (such as the
\emph{differentiation of meanings} and the classification
viewpoint) as the characteristics of novelty detection is
important because they provide new insights into Novelty
theoretically and empirically.

Knowing these properties of the novelty task, we can have two
possible and reasonable directions to deal with the task.

The first possibility, as indicated in section
\ref{sec:complexNV}, is to divide novelty into two subtasks: the
symmetric procedure of identifying PO pairs and the asymmetric
procedure of CO judgment. To improve the novelty detection
performance, this direction would seem to be the most probable and
feasible one, with separate measures (such as error rates of PO
pair and CO relation detection) for the separate subtasks, or with
PSM as a final measure.

If the complexity for evaluating the two separate subtasks is
intolerable, the second possible direction is to treat the novelty
task as a whole, but we should try to find new techniques that
improve only one relation of the two, and use SPSM, SNM and even
precision recall to measure the improvements. One method that
affects both relations can be difficult to evaluate.

Although the experiments in our study were on query-specific
novelty detection datasets, many of the conclusions obtained in
this paper can have a larger generalization to non query-specific
cases. For the novelty task itself, there is still much work to do
following this direction, but we hope this paper as a summary for
one major aspect of the three years' work on Novelty can be a
starting point for those who would like to continue the quest for
efficient novelty detection. For the study of semantics, the
novelty task also provides us a new insight into the
characteristics of meanings: the overlapping relations between
meanings of sentences. In our treatment, unlike previous theories
which consider meanings themselves, we studied meanings of
sentences with the relations between them. We hope this can
inspire new methods and insights to applications dealing with
complicated objects such as the meanings of natural language.

\bibliographystyle{asa}
\bibliography{lzhao05overlap}

\end{document}